# Magnetic symmetry of the plain domain walls in ferro- and ferrimagnets

**B. M. Tanygin[1] and O. V. Tychko[2]***

[1]Kyiv Taras Shevchenko National University, Radiophysics Faculty, Glushkov av.2, build.5, Kyiv, Ukraine, 01022

[1]E-mail: b.m.tanygin@gmail.com

[2]E-mail: pasat@univ.kiev.ua

**Abstract.** Magnetic symmetry of all possible plane domain walls in ferro- and ferrimagnets is considered. Magnetic symmetry classes of non 180 degree (including 0 degree) domain walls are obtained. The domain walls degeneracy is investigated. The symmetry classification is applied for research of all possible plane domain walls in crystals of the hexoctahedral crystallographic class.

## 1. Introduction

The investigation of static and dynamic properties [1,2] of domain walls (DWs) in magnetically ordered media is of considerable interest for the physical understanding of medium behavior and it is also important for applications. For sequential examination of these properties it is necessary to take into account the magnetic symmetry [3,4] of the media. Determination of the DW magnetic symmetry allows

---

*Corresponding author. O.V. Tychko. *Address:* 64 Vladimirskaya str., Taras Shevchenko Kyiv National University, Radiophysics Faculty. 01033 Kyiv, Ukraine. *Tel/fax*: +38-044-526-03-49 *E-mail*: pasat@univ.kiev.ua, a.tychko@mail.ru

to characterize qualitatively some elements of the DW structure and their change. The complete symmetry classification of plane 180 degree DWs ($180^0$-DWs) in magnetically ordered crystals [5] and similar classification of these DWs with Bloch lines in ferromagnets and ferrites [6] were carried out earlier. The plane DWs with width $\delta$ [1,7] exceeding the characteristic size $a$ of a unit magnetic cell were considered. Properties of these DWs in ferro- and ferrimagnets are described by the density of magnetic moment $\boldsymbol{M}$ [8]. Their symmetry can be characterized by the magnetic symmetry classes (MSCs) [9] of a crystal containing a DW [5]. The building of a totality of the MSCs of all possible [1] plane (i.e. DW with $r_0 >> \delta$, where $r_0$ is the curvature radius of the DW [5]) DWs in ferro- and ferrimagnets is the purpose of this work.

## 2. Domain wall symmetry in the magnetically ordered media

Let $\boldsymbol{m}$ be the unit time-odd axial vector [9] along the magnetization vector $\boldsymbol{M}$: $\boldsymbol{m} = \boldsymbol{M} / M$, where $M$ is the saturation magnetization. Then $\boldsymbol{m}_1$ and $\boldsymbol{m}_2$ are unit time-odd axial vectors along magnetization vectors $\boldsymbol{M}_1$ and $\boldsymbol{M}_2$ in neighboring domains: $\boldsymbol{m}_1 = \boldsymbol{M}_1 / M$, $\boldsymbol{m}_2 = \boldsymbol{M}_2 / M$. The vectors $\boldsymbol{m}_1$ and $\boldsymbol{m}_2$ coincide with different easy magnetization axes (EMA) of the medium. The angle $2\alpha$ between these vectors determines the DW type ($2\alpha$-DW): $2\alpha = \arccos(\boldsymbol{m}_1\boldsymbol{m}_2)$. A unit polar time-even vector $\boldsymbol{n}_W$ indicates the DW plane normal. It is directed from domain with $\boldsymbol{m}_1$ to domain with $\boldsymbol{m}_2$. In order to define the unified co-ordinate system we introduce the vectors $\boldsymbol{a}_1$ and $\boldsymbol{a}_2$ as well as the parameters $b_\Sigma = |[\boldsymbol{n}_W \times \boldsymbol{m}_\Sigma]|$ and $b_\Delta = |[\boldsymbol{n}_W \times \Delta\boldsymbol{m}]|$. The unit vectors of the co-ordicate system $O\tilde{x}\tilde{y}\tilde{z}$ are chosen as $[\boldsymbol{e}_{\tilde{x}}, \boldsymbol{e}_{\tilde{y}}, \boldsymbol{e}_{\tilde{z}}] = [\boldsymbol{a}_2, -\boldsymbol{a}_1, \boldsymbol{n}_W]$. Here the unit vector $\boldsymbol{a}_1$ coincides with the direction of the vector $\Delta\boldsymbol{m} - \boldsymbol{n}_W(\boldsymbol{n}_W\Delta\boldsymbol{m})$ (at $b_\Delta \neq 0$ and $b_\Sigma = 0$) or $[\boldsymbol{a}_2 \times \boldsymbol{n}_W]$ (at $b_\Delta = 0$ or $b_\Sigma \neq 0$). The unit vector $\boldsymbol{a}_2$ coincides with the direction of vector $\boldsymbol{m}_\Sigma - \boldsymbol{n}_W(\boldsymbol{n}_W\boldsymbol{m}_\Sigma)$ (at $b_\Sigma \neq 0$) or $[\boldsymbol{n}_W \times \boldsymbol{a}_1]$ (at $b_\Delta \neq 0$ and $b_\Sigma = 0$) or else with an arbitrary direction in the DW plane ($\boldsymbol{a}_2 \perp \boldsymbol{n}_W$ at $b_\Sigma = b_\Delta = 0$). The time-odd axial vectors $\Delta\boldsymbol{m}$ and $\boldsymbol{m}_\Sigma$ are determined by equalities $\Delta\boldsymbol{m} = \boldsymbol{m}_2 - \boldsymbol{m}_1$ and $\boldsymbol{m}_\Sigma = \boldsymbol{m}_1 + \boldsymbol{m}_2$ respectively.

The MSC $G_k$ (here $k$ is a MSC number) of a $2\alpha$-DW is the magnetic symmetry group including all symmetry transformations (here and hereinafter all translations are considered as unit operations) that do not change the spatial distribution of magnetic moments in the crystal with DW [5]. The above-mentioned group is a subgroup of the magnetic (Shubnikov's) symmetry group of the crystal paramagnetic phase [10]. These transformations do not change DW boundary conditions and can be classified by two types [5]. The first type transformations $g^{(1)}$ do not change the directions of the vectors $\boldsymbol{m}_1$, $\boldsymbol{m}_2$ and $\boldsymbol{n}_W$: $g^{(1)}\boldsymbol{n}_W = \boldsymbol{n}_W$, $g^{(1)}\boldsymbol{m}_1 = \boldsymbol{m}_1$, $g^{(1)}\boldsymbol{m}_2 = \boldsymbol{m}_2$. The second type transformations $g^{(2)}$ change these directions: $g^{(2)}\boldsymbol{n}_W = -\boldsymbol{n}_W$, $g^{(2)}\boldsymbol{m}_1 = \boldsymbol{m}_2$, $g^{(2)}\boldsymbol{m}_2 = \boldsymbol{m}_1$. In conformity with the terminology of [6] the MSC $G_B$ of DW boundary conditions is the totality of all transformations of the magnetic symmetry group of the crystal paramagnetic phase that satisfy the mentioned six conditions. It is the MSC of the maximum possible symmetry of a $2\alpha$-DW in the given crystal for a particular mutual orientation of the vectors $\boldsymbol{m}_1$, $\boldsymbol{m}_2$ and $\boldsymbol{n}_W$. The other possible MSCs $G_k$ of a $2\alpha$-DW with fixed directions of the vectors $\boldsymbol{m}_1$, $\boldsymbol{m}_2$ and $\boldsymbol{n}_W$ result by enumeration of the subgroups of $G_B$ : $G_k \subseteq G_B \subset G_P$, where $G_P$ is the MSC of the crystal paramagnetic phase. The mutual orientation of the vectors $\boldsymbol{m}_1$, $\boldsymbol{m}_2$ and $\boldsymbol{n}_W$ is determined by the set of parameters $a_\Sigma = (\boldsymbol{n}_W \boldsymbol{m}_\Sigma)$, $a_\Delta = (\boldsymbol{n}_W \Delta\boldsymbol{m})$, $a_C = (\boldsymbol{n}_W \boldsymbol{m}_C)$, $b_\Sigma$ and $b_\Delta$, where time-even axial vector $\boldsymbol{m}_C$ is determined by equality $\boldsymbol{m}_C = [\boldsymbol{m}_1 \times \boldsymbol{m}_2]$.

The possible MSCs $G_k$ ($1 \le k \le 42$) of $180^0$-DWs were found earlier [5]. All possible MSCs of $2\alpha$-DWs with $2\alpha \ne 180^0$ are presented in table 1.

For a certain $2\alpha$-DW the different MSCs are different groups of magnetic point symmetry transformations. Their representations [11,15] are written in the co-ordinate system $O\tilde{x}\tilde{y}\tilde{z}$. All represented MSCs are not interrelated by a rotation over an arbitrary angle around $\boldsymbol{n}_W$. Also the above-mentioned MSCs are not reduced with each other by unit vectors transformation $\boldsymbol{a}_1 \leftrightarrow \boldsymbol{a}_2$.

The possible transformations $g^{(1)}$ or $g^{(2)}$ (column "Symmetry elements" of table 1) of $2\alpha$-DWs with $2\alpha \ne 180^0$ are rotations around two-fold symmetry axes $2_n$, $2'_n$ or $2_1$, $2'_1$ or else $2_2$, $2'_2$ that are collinear with the unit vectors $\boldsymbol{n}_W$ or $\boldsymbol{a}_1$ or else $\boldsymbol{a}_2$, respectively, reflections in planes $\bar{2}_n$, $\bar{2}'_n$ or $\bar{2}'_1$ or else $\bar{2}_2$, $\bar{2}'_2$ that are normal to the above mentioned vectors, respectively, rotations around three-, four-, six-fold symmetry axes $3_n$, $4_n$, $6_n$ that are collinear with the vector $\boldsymbol{n}_W$, rotations around three-, four-, six-fold inversion axes $\bar{3}_n$, $\bar{4}_n$, $\bar{6}_n$ that are collinear with the vector $\boldsymbol{n}_W$, inversion in the symmetry center $\bar{1}$ and identity (symmetry element 1). Here an accent at symmetry elements means a simultaneous use of the time reversal operation $R$ [9]. For MSCs with $24 \leq k \leq 39$ and $52 \leq k \leq 64$ only generative symmetry elements [11] are represented in table 1.

There is a correspondence between MSCs of 180°-DWs (i.e. at $\boldsymbol{m}_1 = -\boldsymbol{m}_2$ [1]), 0°-DWs (i.e. at $\boldsymbol{m}_1 = \boldsymbol{m}_2$ [13]) and $2\alpha$-DWs with non-collinear orientation of vectors $\boldsymbol{m}_1$ and $\boldsymbol{m}_2$ [1] (hereinafter the last DWs will be marked as $2\alpha'$-DWs). The above mentioned determinations of criterions for transformations $g^{(1)}$ and $g^{(2)}$ can be represented in another identical form: $g^{(1)}\boldsymbol{n}_W = \boldsymbol{n}_W$, $g^{(1)}\boldsymbol{m}_\Sigma = \boldsymbol{m}_\Sigma$, $g^{(1)}\Delta\boldsymbol{m} = \Delta\boldsymbol{m}$ and $g^{(2)}\boldsymbol{n}_W = -\boldsymbol{n}_W$, $g^{(2)}\boldsymbol{m}_\Sigma = \boldsymbol{m}_\Sigma$, $g^{(2)}\Delta\boldsymbol{m} = -\Delta\boldsymbol{m}$. These criterions restrict an ensemble of MSCs symmetry transformations for an arbitrary $2\alpha'$-DW. We have $\Delta\boldsymbol{m} = 0$ and $\boldsymbol{m}_\Sigma = 0$ for 0°- and 180°-DWs, respectively. A pair from the above mentioned criterions does not restrict the MSCs symmetry transformations of 0°- or 180°-DWs. Therefore the magnetic symmetry of $2\alpha'$-DWs does not exceed the magnetic symmetry of 0°- and 180°-DWs generically. The MSCs of 180°-DWs are the MSCs of $2\alpha'$-DWs if their transformations do not break the symmetry of the vector $\boldsymbol{m}_\Sigma$ of the $2\alpha'$-DW (i.e. these MSCs must be subgroup of the group $\infty$/mm′m′, where the infinite-fold symmetry axis is collinear with the vector $\boldsymbol{m}_\Sigma$).

There is an analogy between MSCs of 180°- and 0°-DWs: their transformations $g^{(1)}$ are the same since they belong to a subgroup of axial time-odd vector symmetry group (MSC $\infty$/mm′m′), where the infinite-fold symmetry axis is collinear with $\Delta\boldsymbol{m}$ or $\boldsymbol{m}_\Sigma$ for 180°- or 0°-DWs, respectively. Therefore if MSCs consist of the transformations $g^{(1)}$ only then these MSCs are common for 180°- and 0°-DWs. They are marked with sign “-” in column “DW center” of table 1. A conversion of MSC of 180°-DW into MSC of 0°-DW is simply a change of the criterion $g^{(2)}\Delta\boldsymbol{m} = -\Delta\boldsymbol{m}$ by the criterion $g^{(2)}\boldsymbol{m}_{\Sigma} = \boldsymbol{m}_{\Sigma}$. The transformations of corresponding MSCs of these $2\alpha$-DWs are different by the substitution $g^{(2)} \to g^{(2)} \cdot R$ only. Therefore, if a pair of MSCs of 180°-DWs and a pair of MSCs of 0°-DWs is connected by the above-mentioned substitution, then these MSCs are common for 180°- and 0°-DWs.

As a result the lists of MSCs of 0°-, 180°- and $2\alpha'$-DWs are intersected in general. Total number of MSCs of a $2\alpha$-DW with arbitrary $2\alpha$ value (including $2\alpha$=180$^0$) in ferro- and ferrimagnets is equal to 64. General enumeration of MSCs of 180°-DWs contains 42 MSCs: $1 \le k \le 42$ [5]. This enumeration holds also for MSCs of $2\alpha$-DW with $2\alpha \ne 180°$ (MSC numbers are bold type in column "MSC number *k*" of table 1) . There are 10 MSCs of $2\alpha'$-DWs: $7 \le k \le 13$ and $16 \le k \le 18$. The general list of MSCs of 0°-DWs includes all 42 MSCs of table 1: *k*=2, $6 \le k \le 13$, $16 \le k \le 19$, *k*=22, 24, 26, 30, 32, 37, 39 and $43 \le k \le 64$.

### 3. Domain wall structure

The $2\alpha$-DWs with $\delta >> a$ in ferro- and ferrimagnets are described by the macroscopic density of magnetic moment $\boldsymbol{M}(\tilde{z})$ [5]. The transformations $g^{(1)}$ and $g^{(2)}$ ($g^{(1)} \in G_k$; $g^{(2)} \in G_k$) impose restrictions on the kind of coordinate dependence of $\boldsymbol{m}(\tilde{z})$ components ($\boldsymbol{m}(\tilde{z}) = \boldsymbol{m}_{\tilde{x}}(\tilde{z}) + \boldsymbol{m}_{\tilde{y}}(\tilde{z}) + \boldsymbol{m}_{\tilde{z}}(\tilde{z})$) in the DW volume and allow to find this dependence [5]. For the determination of the kind of coordinate dependence of $\boldsymbol{m}(\tilde{z})$ component of 0°- and $2\alpha'$-DWs for each MSC (column "Coordinate dependences of $\boldsymbol{m}(\tilde{z})$ components" in table 1) the next rules are used: a) if an axial time-odd vector along unit vectors $\boldsymbol{e}_r$ ($r \equiv \tilde{x}, \tilde{y}$ or $\tilde{z}$) is not an invariant of the transformation $g^{(1)}$ then there is no component $\boldsymbol{m}_r(\tilde{z})$ (figure (-) in column "Coordinate dependences of $\boldsymbol{m}(\tilde{z})$ components" of table 1); b) if the axial time-odd vector along $\boldsymbol{e}_r$ is inverted by the transformation $g^{(2)}$ then the component $\boldsymbol{m}_r(\tilde{z})$ is an odd (A) function of coordinate $\tilde{z}$; c) if the axial time-odd vector along $\boldsymbol{e}_r$ is an invariant of the transformation $g^{(2)}$ then $\boldsymbol{m}_r(\tilde{z})$ is an even (S) function of coordinate $\tilde{z}$; d) if the axial time-odd vector along $\boldsymbol{e}_r$ is an invariant of the transformation $g^{(1)}$ then transformation $g^{(1)}$ does not restrict the kind of function $\boldsymbol{m}_r(\tilde{z})$ (A,S).

If the MCS of a $2\alpha$-DW includes transformations that transpose adjacent magnetic domains then this DW has a center of symmetry [5]. These MSCs enclose the symmetry transformations $g^{(2)}$. They are marked by coordinate $\tilde{z}=0$ in column “DW center” of table 1.

As in the case of $180^0$-DWs [5], the $0^0$- DWs can be pulsating (i.e. DW with collinear directions of vectors $\boldsymbol{M}$ and $|\boldsymbol{M}| \neq \text{const}$ in its volume [5]) DWs. The MSCs with $k$=2, 6, 19-45, 49-64 describe symmetry of pulsating DWs only. In contrast with 180°- and 0°-DWs there are no pulsating DWs among the $2\alpha'$-DWs, since $2\alpha'$-DWs require the presence of two “nonzero” $\boldsymbol{m}(\tilde{z})$ components. The $2\alpha'$-DWs are rotary (i.e. DW with $|\boldsymbol{M}|$=const in its volume) or semi-rotary [5] DWs only. Among rotary or semi-rotary DWs there are DWs with only Bloch (i.e DWs with $\boldsymbol{M}\boldsymbol{n}_W$=const) [1,14] ($k$=7, 8 or 46) and only Neel (i.e. DWs with **m** rotation in the plane containing $\boldsymbol{n}_W$) [1,15] ($k$=9, 12, 17 or 47) laws of **m** rotation in their volume.

Crystal magnetic ordering is accompanied by phase transition and change of crystal magnetic symmetry [3]. In a magnetically ordered crystal $q_k$-multiply degenerate $2\alpha$-DWs with fixed $2\alpha$ can be obtained [6], where $q_k = \text{ord}(G_P)/\text{ord}(G_k)$. Functions $\text{ord}(G_P)$ and $\text{ord}(G_k)$ give the order [11] of the magnetic point group of the crystal paramagnetic phase [9,10] and of a $2\alpha$-DW in this crystal, respectively. These $2\alpha$-DWs have the same energy but different structures (magnetization distribution, plane orientation, etc.). The minimum value of $q_k$ is 2 in accordance with the invariance of energy for time reversal operation *R*.

At representation of the $G_P$ as the totality of $G_k$ (with fixed value $k$ and different symmetry elements orientations) the lost transformations (members of adjacent classes) $g^l$ [6,12] interrelate the above mentioned $q_k$-multiply degenerate $2\alpha$-DWs (i.e. $g^l$ operation converts an one of such $2\alpha$-DWs into another).

The degeneracy $q_k$ of a $2\alpha$-DW can be written in the form $q_k = q_B q'_k$ ($q'_k \le q_k$), where $q'_k = \text{ord}(G_B)/\text{ord}(G_k)$ is the number of equal-energy $2\alpha$-DWs with fixed boundary conditions, $q_B = \text{ord}(G_P)/\text{ord}(G_B)$ is the number of possible boundary conditions. Here $\text{ord}(G_B)$ is the order of the point group of the maximum magnetic symmetry of the $2\alpha$-DW in the given crystal.

The $2\alpha$-DWs of MSC $G_{16}$ (MSC 1) have the maximum degeneracy $q_k$. For 180°- and $2\alpha'$-DWs it is equal to 16 (crystallographic class mmm), 48 (crystallographic class 6/mmm) and 96 (crystallographic class m3m) in crystals of lower, medium and higher symmetry singonies (in conformity with terminology of [11]), respectively. The 0°-DWs are formed in spatially inhomogeneous media [13]. Conditions of occurrence and existence of such DWs demand to take into account medium peculiarities.

### 4. Magnetic symmetry classes of domain walls in hexoctahedral crystals

As an example let's consider MSCs of all possible DWs in magnetically ordered crystals of hexoctahedral class (crystallographic point symmetry group $\text{m}3\text{m}$ in the paramagnetic phase [3]) . This class is assumed to exhibit the largest variety of possible DWs. Furthermore it encompasses widely investigated and used magnetic media (all cubic symmetry metals, specifically iron and nickel [6], magnetic oxides, specifically ferrites with structures of spinel [4] and garnet [16], perovskite, magnetite and others).

The magnetic anisotropy (MA) energy $e_K$ is the invariant of the initial paramagnetic phase of crystal. For the $\text{m}3\text{m}$ crystal this energy is given by $e_K(\alpha_1, \alpha_2, \alpha_3) = K_1\, s + K_2\, p + K_3\, s^2 + K_4\, sp + ...$, where $K_1$, $K_2$, $K_3$ and $K_4$ are first, second, third and fourth MA constants, $s = \alpha_1^2\alpha_2^2 + \alpha_2^2\alpha_3^2 + \alpha_1^2\alpha_3^2$, $p = \alpha_1^2\alpha_2^2\alpha_3^2$, $\alpha_1$, $\alpha_2$ and $\alpha_3$ are the direction cosines of $\boldsymbol{m}$ [16]. The absolute minimum of this energy corresponds to EMAs. Signs of MA constants and relation between their values determine EMAs directions. In the framework of the ($K_1$, $K_2$, $K_3$) approximation the EMAs directions can coincide with both high-symmetric and low-symmetric crystallographic directions [17]. In the framework of the two-constant ($K_1, K_2$) approximation the EMA directions can coincide only with high-symmetric <111> or <110> or else <100> like crystallographic directions at $K_1 \leq -K_2/3$ or $0 \geq K_1 \geq -K_2/2$ or else $K_1 \geq 0$ respectively [1,18]. At that $71^0$-, $109^0$- and $180^0$-DWs or $60^0$-, $90^0$-, $120^0$- and $180^0$-DWs or else $90^0$- and $180^0$-DWs are realized in a $m3m$ crystal, respectively [1]. The MSCs and degeneracy $q_B$ of a $2\alpha$-DW boundary conditions with $2\alpha > 90°$ and $2\alpha \leq 90°$ are presented in tables 2 and 3 respectively.

The earlier obtained MSCs of merely 180°-DWs (bold type numbers in table 2) include elements [5]: *k*=1 - $(1, \bar{2}_1, 2_2, \bar{2}_n) \times (1, \bar{1}')$; *k*=4 - $(1, \bar{1}', 2'_1, \bar{2}_1)$; *k*=5 - $(1, \bar{1}', 2'_n, \bar{2}_n)$; *k*=14 - $(1, \bar{1}', 2_2, \bar{2}'_2)$; *k*=15 - $(1, \bar{1}')$; *k*=23 - $(1, 2_1, 2_2, 2_n) \times (1, \bar{1}')$; *k*=29 - $(\bar{3}'_n, \bar{2}'_1)$; *k*=34 - $(4_n, \bar{2}'_1, \bar{2}'_n)$.

Only generative symmetry elements are presented for *k*=29 and 34. Other MSCs of tables 2 and 3 are presented in table 1. In these tables the DW plane orientation is assigned by different Miller indexes *h,k,l*>1. A simultaneous change on negative and/or cyclic permutation of all indexes doesn't change MSCs.

There are no common MSCs of maximum symmetrical 180°- and $2\alpha'$-DWs in the m3m crystal. It is connected with the presence of the $\bar{1}'$ transformation ($2\alpha'$-DW vector $\boldsymbol{m}_\Sigma$ is changed by this transformation) in the MSCs of such 180°-DW.

**5. Conclusions**

The full magnetic symmetry classification of all possible domain walls in ferro- and ferrimagnet crystals includes 64 magnetic symmetry classes: 42 classes of $0^0$- DWs, 10 classes of $2\alpha$-DWs with $0^0 < 2\alpha < 180^0$ and 42 classes of $180^0$-DWs. Lists of magnetic symmetry classes of all above mentioned types of DWs are intersected in general case.

$0^0$- DWs can be pulsating, rotary or semi-rotary DWs. The $2\alpha$-DWs with $0^0 < 2\alpha < 180^0$ are rotary or semi-rotary DWs only. Among rotary or semi-rotary DWs there are DWs with Bloch or Neel laws of magnetization rotation in their volume. Pulsating, rotary or semi-rotary DWs can have a center of symmetry in their volume.

All possible $180^0$- and $2\alpha$-DWs with $0^0 < 2\alpha < 180^0$ have even degeneracy (its value is between 2 and 96 in general case).

Magnetic symmetry classes of maximum symmetrical 180°-DWs do not meet with such classes of $2\alpha$-DWs with $0^0 < 2\alpha < 180^0$ in a m3m crystal.

**References**

[1] A. Hubert, Theorie der Domanenwande in Geordneten Medielen (Theory of Domain Walls in Ordered Media), Springer, Berlin, Heidelberg, New York, 1974

A Hubert and R. Shafer, Magnetic Domains. The Analysis of Magnetic Microstructures, Springer, Berlin, 1998

[2] V. Bokov and V. Volkov, Physics of the Solid State 50 (2008)198

[3] L. Shuvalov, Sov. Phys. Crystallogr. 4(1959)399

[4] L. Shuvalov, Modern Crystallography IV : Physical Properties of Crystals, Springer, Berlin, 1988

[5] V. Baryakhtar, V. Lvov and D. Yablonsky, JETP 87(1984)1863

[6] V. Baryakhtar, E. Krotenko and D. Yablonsky, JETP 91(1986)921

[7] B. Lilley, Phil.Mag. 41(1950)792

[8] A. Andreev and V. Marchenko, JETP 70(1976)1522

[9] L. Landau, E. Lifshitz and L. Pitaevskii, Course of Theoretical Physics, vol.8. Electrodynamics of Continuous Media, Pergamon Press, London, 1984

[10] V. A. Kopcik, Xubnikovskie Gruppy: Spravoqnik po simmetrii i fiziqeskim svostvam. kristalliqeskih struktur [Shubnikov's groups: Handbook on the symmetry and physical properties of crystalline structures, in Russian], Izdatel'stvo Moskovskogo Universiteta, Moscow, 1966

A.V. Shubnikov and N.V. Belov, Colored symmetry, Pergamon Press, London, 1964

B. Tavger and V. Zaitzev, JETP 3(1956)430

[11] B. Vanshtein, Modern Crystallography 1: Symmetry of Crystals, Methods of Structural Crystallography, Springer, Berlin, 1994

[12] E. Wigner, Group Theory and its Application to the Quantum Mechanics of Atomic Spectra, Academic Press, New York, 1959

[13] L. Heyderman, H. Niedoba, H. Gupta and I. Puchalska, J. Magn. Magn. Mater 96(1991)125.

R. Vakhitov, A Yumaguzin, J. Magn. Magn. Mater. 215-216(2000)52

[14] L.Landau and E.Lifshitz, Sov.Phys. 8(1935)153

[15] L. Neel, Compt.rend. 2419(1955)533

[16] A. Paoletti, Physics of Magnetic Garnets, Esevier, Amsterdam, 1978

[17] U. Atzmony and M. Dariel, Phys. Rev. B13(1976)4006

[18] K.P. Belov, A.K. Zvezdin, R.Z. Levitin, A.S. Markosyan, B.V. Mill’, A.A. Mukhin and A.P.Perov, JETP 41(1975)590

**Table 1.** Magnetic symmetry classes of the plane $2\alpha$-DWs with $2\alpha \neq 180°$.

| MSC numb. $k$ | Mutual orientations of the vectors $\boldsymbol{m}_1$, $\boldsymbol{m}_2$ and $\boldsymbol{n}_W$ | Symmetry elements | Coordinate dependences of $\boldsymbol{m}(\tilde{z})$ components | | | DW center | International MSC symbol |
|---|---|---|---|---|---|---|---|
| | | | $\boldsymbol{m}_{\tilde{y}}(\tilde{z})$ | $\boldsymbol{m}_{\tilde{x}}(\tilde{z})$ | $\boldsymbol{m}_{\tilde{z}}(\tilde{z})$ | | |
| **2** | $a_\Delta = b_\Delta = a_\Sigma = 0$ | $1, \bar{2}'_1, \bar{2}_2, 2'_n$ | (-) | (A,S) | (-) | - | $\mathrm{mm}'\,2'$ |
| **6** | $a_\Sigma = a_\Delta = a_C = 0$ | $1, \bar{2}_2$ | (-) | (A,S) | (-) | - | m |
| **7** | $a_\Sigma = a_\Delta = 0$ | $1,\ 2'_1, 2_2, 2'_n$ | (A) | (S) | (-) | $\tilde{z} = 0$ | $2\,2'\,2'$ |
| **8** | $a_\Sigma = a_\Delta = 0$ | $1,\ 2'_n$ | (A,S) | (A,S) | (-) | - | $2'$ |
| **9** | $a_C = a_\Delta = b_\Sigma = 0$ | $1,\ 2'_1, \bar{2}'_2, \bar{2}_n$ | (A) | (-) | (S) | $\tilde{z} = 0$ | $\mathrm{m\,m}'\,2'$ |
| **10** | $a_\Delta = 0$ | $1,\ 2'_1$ | (A) | (S) | (S) | $\tilde{z} = 0$ | $2'$ |
| **11** | $a_C = a_\Delta = b_\Sigma = 0$ | $1,\ \bar{2}_n$ | (A) | (A) | (S) | $\tilde{z} = 0$ | m |
| **12** | $a_C = 0$ | $1, \bar{2}'_1$ | (-) | (A,S) | (A,S) | - | $\mathrm{m}'$ |
| **13** | $a_\Sigma = 0$ | $1,\ 2_2$ | (A) | (S) | (A) | $\tilde{z} = 0$ | 2 |
| **16** | Arbitrary | 1 | (A,S) | (A,S) | (A,S) | - | 1 |
| **17** | $a_C = a_\Sigma = b_\Delta = 0$ | $1,\ \bar{2}'_1, 2_2, \bar{2}'_n$ | (-) | (S) | (A) | $\tilde{z} = 0$ | $\mathrm{m}'\,\mathrm{m}'\,2$ |
| **18** | $a_C = a_\Sigma = b_\Delta = 0$ | $1,\ \bar{2}'_n$ | (S) | (S) | (A) | $\tilde{z} = 0$ | $\mathrm{m}'$ |
| **19** | $b_\Delta = b_\Sigma = 0$ | $1,\ 2_n$ | (-) | (-) | (A,S) | - | 2 |
| **22** | $b_\Delta = b_\Sigma = 0$ | $1, \bar{2}'_1, \bar{2}'_2, 2_n$ | (-) | (-) | (A,S) | - | $\mathrm{m}'\,\mathrm{m}'\,2$ |
| **24** | $b_\Delta = b_\Sigma = 0$ | $3_n$ | (-) | (-) | (A,S) | - | 3 |
| **26** | $b_\Delta = b_\Sigma = 0$ | $3_n, \bar{2}'_1$ | (-) | (-) | (A,S) | - | $3\mathrm{m}'$ |
| **30** | $b_\Delta = b_\Sigma = 0$ | $4_n$ | (-) | (-) | (A,S) | - | 4 |
| **32** | $b_\Delta = b_\Sigma = 0$ | $4_n, \bar{2}'_1$ | (-) | (-) | (A,S) | - | $4\mathrm{m}'\mathrm{m}'$ |
| **37** | $b_\Delta = b_\Sigma = 0$ | $6_n$ | (-) | (-) | (A,S) | - | 6 |
| **39** | $b_\Delta = b_\Sigma = 0$ | $6_n, \bar{2}'_1$ | (-) | (-) | (A,S) | - | $6\mathrm{m}'\mathrm{m}'$ |
| 43 | $a_\Delta = b_\Delta = a_\Sigma = 0$ | $(1, 2'_1, \bar{2}_2, \bar{2}'_n) \times (1, \bar{1})$ | (-) | (S) | (-) | $\tilde{z} = 0$ | $\mathrm{mm}'\mathrm{m}'$ |
| 44 | $a_\Delta = b_\Delta = a_\Sigma = 0$ | $1, 2'_1, \bar{2}_2, \bar{2}'_n$ | (-) | (S) | (-) | $\tilde{z} = 0$ | $\mathrm{mm}'\,2'$ |
| 45 | $a_\Delta = b_\Delta = a_\Sigma = 0$ | $1, \bar{1}, 2_2, \bar{2}_2$ | (-) | (S) | (-) | $\tilde{z} = 0$ | 2/m |
| 46 | $a_\Delta = b_\Delta = a_\Sigma = 0$ | $1, \bar{1}, 2'_n, \bar{2}'_n$ | (S) | (S) | (-) | $\tilde{z} = 0$ | $2'/\mathrm{m}'$ |
| 47 | $a_\Delta = b_\Delta = 0$ | $1, \bar{1}, 2'_1, \bar{2}'_1$ | (-) | (S) | (S) | $\tilde{z} = 0$ | $2'/\mathrm{m}'$ |
| 48 | $a_\Delta = b_\Delta = 0$ | $1, \bar{1}$ | (S) | (S) | (S) | $\tilde{z} = 0$ | $\bar{1}$ |

**Table 1.** Magnetic symmetry classes of the plane $2\alpha$-DWs with $2\alpha \neq 180°$ (continue).

| MSC numb. $k$ | Mutual orientations of the vectors $\boldsymbol{m}_1$, $\boldsymbol{m}_2$ and $\boldsymbol{n}_W$ | Symmetry elements | Coordinate dependences of $\boldsymbol{m}(\tilde{z})$ components | | | DW center | International MSC symbol |
|---|---|---|---|---|---|---|---|
| | | | $\boldsymbol{m}_{\tilde{y}}(\tilde{z})$ | $\boldsymbol{m}_{\tilde{x}}(\tilde{z})$ | $\boldsymbol{m}_{\tilde{z}}(\tilde{z})$ | | |
| 49 | $a_\Delta = b_\Delta = b_\Sigma = 0$ | $1, \bar{1}, 2_n, \bar{2}_n$ | (-) | (-) | (S) | $\tilde{z} = 0$ | 2/m |
| 50 | $a_\Delta = b_\Delta = b_\Sigma = 0$ | $1, 2'_1, 2'_2, 2_n$ | (-) | (-) | (S) | $\tilde{z} = 0$ | $2\,2'\,2'$ |
| 51 | $a_\Delta = b_\Delta = b_\Sigma = 0$ | $(1, 2'_1, 2'_2, 2_n) \times (1, \bar{1})$ | (-) | (-) | (S) | $\tilde{z} = 0$ | mm′m′ |
| 52 | $a_\Delta = b_\Delta = b_\Sigma = 0$ | $\bar{6}_n$ | (-) | (-) | (S) | $\tilde{z} = 0$ | $\bar{6}$ |
| 53 | $a_\Delta = b_\Delta = b_\Sigma = 0$ | $3_n, 2'_1$ | (-) | (-) | (S) | $\tilde{z} = 0$ | $32'$ |
| 54 | $a_\Delta = b_\Delta = b_\Sigma = 0$ | $\bar{6}_n, 2'_1$ | (-) | (-) | (S) | $\tilde{z} = 0$ | $\bar{6}$m′2′ |
| 55 | $a_\Delta = b_\Delta = b_\Sigma = 0$ | $\bar{3}_n, \bar{2}'_1$ | (-) | (-) | (S) | $\tilde{z} = 0$ | $\bar{3}$m′ |
| 56 | $a_\Delta = b_\Delta = b_\Sigma = 0$ | $4_n, \bar{2}_n$ | (-) | (-) | (S) | $\tilde{z} = 0$ | 4/m |
| 57 | $a_\Delta = b_\Delta = b_\Sigma = 0$ | $4_n, 2'_1$ | (-) | (-) | (S) | $\tilde{z} = 0$ | $42'2'$ |
| 58 | $a_\Delta = b_\Delta = b_\Sigma = 0$ | $4_n, \bar{2}'_1, \bar{2}_n$ | (-) | (-) | (S) | $\tilde{z} = 0$ | 4/mm′m′ |
| 59 | $a_\Delta = b_\Delta = b_\Sigma = 0$ | $\bar{4}_n$ | (-) | (-) | (S) | $\tilde{z} = 0$ | $\bar{4}$ |
| 60 | $a_\Delta = b_\Delta = b_\Sigma = 0$ | $\bar{4}_n, 2'_1$ | (-) | (-) | (S) | $\tilde{z} = 0$ | $\bar{4}$2′m′ |
| 61 | $a_\Delta = b_\Delta = b_\Sigma = 0$ | $6_n, \bar{2}_n$ | (-) | (-) | (S) | $\tilde{z} = 0$ | 6/m |
| 62 | $a_\Delta = b_\Delta = b_\Sigma = 0$ | $6_n, 2'_1$ | (-) | (-) | (S) | $\tilde{z} = 0$ | $62'2'$ |
| 63 | $a_\Delta = b_\Delta = b_\Sigma = 0$ | $6_n, \bar{2}'_1, \bar{2}_n$ | (-) | (-) | (S) | $\tilde{z} = 0$ | 6/mm′m′ |
| 64 | $a_\Delta = b_\Delta = b_\Sigma = 0$ | $\bar{3}_n$ | (-) | (-) | (S) | $\tilde{z} = 0$ | $\bar{3}$ |

**Table.2.** Number *k* (degeneracy $q_B$) of MSC of boundary conditions of arbitrary oriented plane $2\alpha$-DW ($2\alpha > 90°$) in the cubic $m\bar{3}m$ crystals at selected domain magnetization directions.

| DW plane | $2\alpha$-DW boundary conditions | | | | |
|---|---|---|---|---|---|
| | 180°-DW $[100],[\bar{1}00]$ | 180°-DW $[110],[\bar{1}\bar{1}0]$ | 180°-DW $[111],[\bar{1}\bar{1}\bar{1}]$ | 120°-DW $[110],[0\bar{1}1]$ | 109°-DW $[111],[1\bar{1}\bar{1}]$ |
| (100) | **34** (6) | **14** (24) | **14** (24) | 16 (96) | 9 (24) |
| (010) | **1** (12) | **14** (24) | **14** (24) | 13 (48) | 13 (48) |
| (001) | **1** (12) | **1** (12) | **14** (24) | 16 (96) | 13 (48) |
| (111) | **14** (24) | **14** (24) | **29** (8) | 16 (96) | 12 (48) |
| $(\bar{1}11)$ | **14** (24) | **4** (24) | **14** (24) | 13 (48) | 12 (48) |
| $(1\bar{1}1)$ | **14** (24) | **4** (24) | **14** (24) | 16 (96) | 10 (48) |
| $(11\bar{1})$ | **14** (24) | **14** (24) | **14** (24) | 13 (48) | 10 (48) |
| (110) | **14** (24) | **23** (12) | **14** (24) | 16 (96) | 16 (96) |
| (101) | **14** (24) | **15** (48) | **14** (24) | 11 (48) | 16 (96) |
| (011) | **1** (12) | **15** (48) | **14** (24) | 16 (96) | 17 (24) |
| $(\bar{1}10)$ | **14** (24) | **1** (12) | **5** (24) | 16 (96) | 16 (96) |
| $(\bar{1}01)$ | **14** (24) | **15** (48) | **5** (24) | 13 (48) | 16 (96) |
| $(0\bar{1}1)$ | **1** (12) | **15** (48) | **5** (24) | 16 (96) | 7 (24) |
| (*hhl*) | **15** (48) | **14** (24) | **14** (24) | 16 (96) | 16 (96) |
| (*hkh*) | **15** (48) | **15** (48) | **14** (24) | 16 (96) | 16 (96) |
| (*hkk*) | **14** (24) | **15** (48) | **14** (24) | 16 (96) | 12 (48) |
| $(\bar{h}hl)$ | **15** (48) | **4** (24) | **15** (48) | 16 (96) | 16 (96) |
| $(\bar{h}kh)$ | **15** (48) | **15** (48) | **15** (48) | 13 (48) | 16 (96) |
| $(h\bar{k}k)$ | **14** (24) | **15** (48) | **15** (48) | 16 (96) | 10 (48) |
| $(hk0),(\bar{h}k0)$ | **14** (24) | **14** (24) | **15** (48) | 16 (96) | 16 (96) |
| $(h0l),(\bar{h}0l)$ | **14** (24) | **15** (48) | **15** (48) | 16 (96) | 16 (96) |
| $(0kl),(0\bar{k}l)$ | **4** (24) | **15** (48) | **15** (48) | 16 (96) | 13 (48) |
| $(hkl),(\bar{h}kl),(h\bar{k}l),(hk\bar{l})$ | **15** (48) | **15** (48) | **15** (48) | 16 (96) | 16 (96) |

**Table.3.** Number *k* (degeneracy $q_B$) of MSC of boundary conditions of arbitrary oriented plane $2\alpha$-DW ($2\alpha \leq 90°$) in the cubic $\mathrm{m}\bar{3}\mathrm{m}$ crystals at selected domain magnetization directions.

| DW plane | $2\alpha$-DW boundary conditions | | | |
|---|---|---|---|---|
| | 90°-DW $[100], [0\bar{1}0]$ | 90°-DW $[110], [1\bar{1}0]$ | 71°-DW $[111], [\bar{1}11]$ | 60°-DW [110], [011] |
| (100) | 12 (48) | 9 (24) | 17 (24) | 16 (96) |
| (010) | 12 (48) | 17 (24) | 10 (48) | 10 (48) |
| (001) | 7 (24) | 7 (24) | 10 (48) | 16 (96) |
| (111) | 13 (48) | 16 (96) | 12 (48) | 10 (48) |
| $(\bar{1}11)$ | 10 (48) | 16 (96) | 12 (48) | 16 (96) |
| $(1\bar{1}1)$ | 10 (48) | 16 (96) | 13 (48) | 10 (48) |
| $(11\bar{1})$ | 13 (48) | 16 (96) | 13 (48) | 16 (96) |
| (110) | 17 (24) | 12 (48) | 16 (96) | 16 (96) |
| (101) | 16 (96) | 10 (48) | 16 (96) | 10 (48) |
| (011) | 16 (96) | 13 (48) | 9 (24) | 16 (96) |
| $(\bar{1}10)$ | 9 (24) | 12 (48) | 16 (96) | 16 (96) |
| $(\bar{1}01)$ | 16 (96) | 10 (48) | 16 (96) | 18 (48) |
| $(0\bar{1}1)$ | 16 (96) | 13 (48) | 7 (24) | 16 (96) |
| $(hhl)$ | 13 (48) | 16 (96) | 16 (96) | 16 (96) |
| $(hkh)$ | 16 (96) | 16 (96) | 16 (96) | 10 (48) |
| $(hkk)$ | 16 (96) | 16 (96) | 12 (48) | 16 (96) |
| $(\bar{h}hl)$ | 10 (48) | 16 (96) | 16 (96) | 16 (96) |
| $(\bar{h}kh)$ | 16 (96) | 16 (96) | 16 (96) | 16 (96) |
| $(h\bar{k}k)$ | 16 (96) | 16 (96) | 13 (48) | 16 (96) |
| $(hk0), (\bar{h}k0)$ | 12 (48) | 12 (48) | 16 (96) | 16 (96) |
| $(h0l), (\bar{h}0l)$ | 16 (96) | 10 (48) | 16 (96) | 16 (96) |
| $(0kl), (0\bar{k}l)$ | 16 (96) | 13 (48) | 10 (48) | 16 (96) |
| $(hkl), (\bar{h}kl), (h\bar{k}l), (hk\bar{l})$ | 16 (96) | 16 (96) | 16 (96) | 16 (96) |